\documentclass[reprint, amsmath,amssymb,aps,prb, superscriptaddress]{revtex4-2}
\usepackage{graphicx}
\usepackage{dcolumn}
\usepackage{bm}

\usepackage{xcolor}
\usepackage{longtable}
\usepackage{colortbl}%
  
\usepackage{booktabs,tabularx}
\usepackage{dcolumn}
\usepackage{hyperref}

\hypersetup{
    colorlinks = true,
    citecolor = {blue},
    linkcolor = {blue},
    urlcolor  = {blue},
}

\usepackage{array}
\newcolumntype{P}[1]{>{\centering\arraybackslash}p{#1}}
\newcolumntype{M}[1]{>{\centering\arraybackslash}m{#1}}

\begin{document}

\ 

\

\

\title{Superconductivity, Kondo physics and magnetic order: Tuning the groundstate in the La$_{1-x}$Ce$_x$FeSiH solid solution through the interplay between $3d$ and $4f$ correlated electrons}

\author{J. Sourd}
\affiliation{University of Bordeaux, CNRS, LOMA  UMR 5798, F-33400 Talence, France}
\affiliation{Univ. Bordeaux, CNRS, Bordeaux INP, ICMCB, UMR 5026, F-33600 Pessac, France}
\affiliation{Hochfeld-Magnetlabor Dresden (HLD-EMFL) and Würzburg-Dresden Cluster of Excellence ct.qmat,
Helmholtz-Zentrum Dresden-Rossendorf, 01328 Dresden, Germany}
\author{B. Vignolle}
\affiliation{Univ. Bordeaux, CNRS, Bordeaux INP, ICMCB, UMR 5026, F-33600 Pessac, France}
\author{E. Gaudin}
\affiliation{Univ. Bordeaux, CNRS, Bordeaux INP, ICMCB, UMR 5026, F-33600 Pessac, France}
\author{S. Burdin}
\affiliation{University of Bordeaux, CNRS, LOMA  UMR 5798, F-33400 Talence, France}
\author{S. Tencé}
\affiliation{Univ. Bordeaux, CNRS, Bordeaux INP, ICMCB, UMR 5026, F-33600 Pessac, France}

\date{\today}\begin{abstract}
We report a study of the La$_{1-x}$Ce$_x$FeSiH solid solution  ($0 \leq x \leq 1$), a family of intermetallic hydrides of ZrCuSiAs-type structure, with space group $P4/nmm$. For low cerium concentrations $x \leq 0.20$, we observe the presence of superconductivity, which originates from the correlated $3d$ electrons of iron. The superconducting regime is progressively suppressed by the cerium substitution. For moderate cerium concentration $0.07 \leq  x \leq 0.50$, we observe evidence of the single-ion Kondo effect and no magnetic phase transition down to 2 K. For $0.07 \leq  x \leq 0.20$, the single-ion Kondo effect coexists with a superconducting ground state at low temperatures. From $x > 0.50$, we observe signatures of Kondo coherence and a heavy Fermi liquid regime at low temperature. Finally, at high cerium concentration $x \geq 0.85$, we observe signatures of magnetic ordering at low temperatures. We discuss our results by introducing temperature scales related to superconductivity, the Kondo effect, and magnetic order, which permits building a rich phase diagram temperature versus cerium content $x$. This shows that using the cerium concentration $x$ as a unique control parameter, we can explore the Kondo entanglement between correlated $3d$ and $4f$ electrons, which suggests an unusual change between the superconducting state related to the $3d$ electrons and the Kondo coherent state involving both $3d$ and $4f$ electrons.
\end{abstract}

\maketitle

\section{\label{sec:level1}Introduction}

The study of compounds containing rare earths and transition metals provide plenty of exotic quantum phases of matter at low temperatures, such as frustrated magnetic orders \cite{vojta2018frustration}, heavy fermions \cite{coleman2001fermiliquids}, unconventional superconductivity \cite{pfleiderer2009superconducting,aoki2022unconventional}, unconventional quantum phase transitions and quantum critical points \cite{gegenwart2002magnetic,paglione2003field},  or multipolar orders \cite{manago2021unusual}. For almost two decades, a special interest has been devoted to compounds of the ZrCuSiAs structure containing iron, such as LaFeAsO \cite{kamihara2008iron} and LaFePO \cite{kamihara2006iron}, showing unconventional iron-based superconductivity \cite{stewart2011superconductivity}. In this context, the LaFeSiH \cite{bernardini2018iron} compound is particularly interesting, being the first member of the iron-based superconductor family, which is based on the FeSi layer. Furthermore, LaFeSiH shows some peculiar properties: superconductivity is observed below $T_c$=11 K without doping, and no net magnetic moment on the iron atom is observed in Mössbauer spectroscopy \cite{hansen2024magnetic}. Superconductivity is also observed by replacing the hydrogen atom by fluorine in LaFeSiF$_x$ \cite{vaney2022topotactic} or by oxygen in LaFeSiO$_x$ \cite{hansen2022superconductivity}.

In iron-based superconductors of the ZrCuSiAs structure, the substitution of non-magnetic lanthanum by magnetic cerium shows some very interesting properties, such as magnetic order in CeFeSiH \cite{sourd2024drastic}, superconductivity induced under pressure in CeFeAsO \cite{mydeen2020electron}, or heavy-fermion behavior in CeFePO \cite{bruning2008cefepo}. Cerium-containing intermetallics such as CeFeSiH are often referred to as Kondo lattices. In Kondo lattices, the interesting physics is generally associated with the magnetic Kondo interaction between the cerium-localized $4f^1$ electron and the itinerant-electron spin density. Depending on the system, the Kondo interaction can lead to a magnetic ordered state of the $4f^1$ electron, as observed in CeFeAsO  \cite{mydeen2020electron}, or to a paramagnetic state resulting from the formation of local singlets between the $4f^1$ and the conduction electrons, as observed in CeFePO \cite{bruning2008cefepo}. This competition is explained through the Doniach argument \cite{doniach1977kondo}. Furthermore, in the study of Kondo systems, the distinction is generally made between a Kondo incoherent regime, where the $4f^1$ electrons at each cerium site interact independently with the conduction electrons, and a Kondo coherent regime, where the $4f^1$ electrons interact collectively with the conduction sea \cite{burdin2000coherence}. The passage from the incoherent to the coherent Kondo regime is realized in Kondo alloys \cite{burdin2013lifshitz,poudel2021phase} such as in the magnetic La$_{1-x}$Ce$_x$Cu$_2$Ge$_2$ \cite{hodovanets2015remarkably} solid solution or in the paramagnetic {L}a$_{1-x}${C}e$_x${N}i$_2${G}e$_2$ \cite{pikul2012single} solid solution. In Kondo alloys, the cerium atom is progressively substituted with a non-magnetic atom such as lanthanum, and one can distinguish the Kondo incoherent from the Kondo coherent regimes thanks to different physical probes such as the resistivity or the specific heat. 

In this paper, we report the study of the La$_{1-x}$Ce$_x$FeSiH solid solution, which interpolates between the iron-based superconductivity of LaFeSiH and the cerium-based Kondo-lattice coherence of CeFeSiH. Our results are given by means of x-ray diffraction, transport and thermodynamic measurements. By introducing temperature scales associated with the different regimes, we build a rich phase diagram temperature versus cerium content $x$. This shows how chemical atomic substitution offers a very special playground to tune the degree of entanglement between the cerium $4f$ electrons and the superconducting iron $3d$ electrons.

\section{Experimental Procedures}\label{sec2}

We synthesized polycrystalline La$_{1-x}$Ce$_x$FeSi samples by arc melting a stoichiometric mixture of pure elements (metal basis purity $> 99.9\%$ for iron and silicon, and $>99.8\%$ for lanthanum and cerium) under a high-purity argon atmosphere. We melted the sample several times in order to ensure uniformity. Total mass loss was less than 1$\%$. We performed annealing of the powdered sample in evacuated quartz tubes for one week at 950°C. The powdering was necessary to maximally reduce the presence of La$_{1-x}$Ce$_x$Fe$_2$Si$_2$ secondary phase. We realized hydrogen insertion on the annealed powder, heated under dynamical vacuum at 250°C and then exposed for 4 hours under 10 bars of hydrogen gas at the same temperature.

We performed Powder x-ray diffraction (PXRD) with the use of a PANalytical X'pert Pro diffractometer (Cu-K$\alpha$ radiation) at room temperature, in the range $10$°$ < 2\theta < 100$°. We carried out data analysis using the Rietveld method using the Fullprof program package. We realized magnetization measurements in a SQUID device (Quantum Design MPMS-XL) in the range 1.8-300K, using fields up to 5 T. Before preparing samples for magnetic measurements, we put the powder in front of a magnet at room temperature in order to extract some elemental iron residuals (estimated $< 0.5$ wt $\%$ from SQUID measurements without the magnetic sorting procedure). We conducted transport measurements on compacted powder pellets (compactness $> 80\%$) of 20 mg for electrical resistance and heat capacity and 60 mg for thermoelectric power. We realized electrical resistance measurements in the range 2-300K using the standard four-probe method with silver paint contacts and a current intensity of 10 mA on a Physical Property Measurement System (PPMS) setup. We performed heat-capacity measurements with the relaxation method with a Quantum Design PPMS system and using a two-tau model analysis. We carried out thermoelectric power measurements using the dynamical method described in Ref. \onlinecite{dordor1980dispositif}.

\section{Experimental Results}\label{sec3}

\subsection{Structural properties}

\begin{figure}
    \centering
    \includegraphics[width=\linewidth]{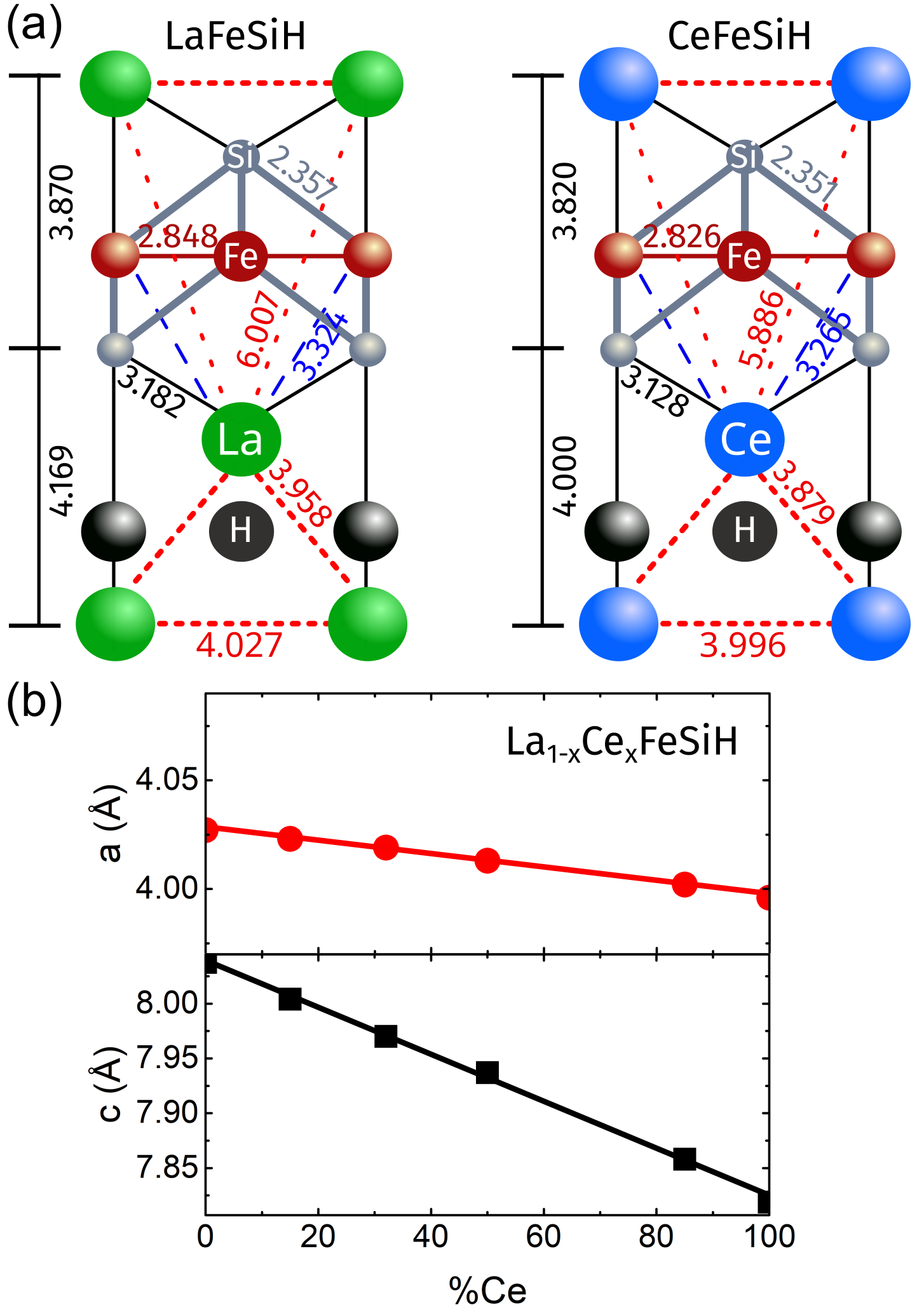}
    \caption{ (a) Selected interatomic distances in LaFeSiH and CeFeSiH (in $\textup{\AA}$). (b) Evolution of lattice parameters in La$_{1-x}$Ce$_x$FeSiH solid solution at room temperature with respect to cerium concentration.}\label{fig1} 
\end{figure}
The PXRD results indicate a CeFeSi-type structure of symmetry $P4/nmm$ for all compositions of the La$_{1-x}$Ce$_x$FeSi system, as well as for all compositions of the La$_{1-x}$Ce$_x$FeSiH system. The La$_{1-x}$Ce$_x$FeSi parent phase structure contains layers of empty rare-earth tetrahedra alternating with layers of Si tetrahedra filled by Fe atoms. We assume that, in La$_{1-x}$Ce$_x$FeSiH hydrides, hydrogen atoms are located in the rare-earth tetrahedra with an occupation of 100$\%$ (Wyckoff position 2b) based on previous neutron diffraction studies performed on hydrogenated/deuterated isostructural compounds, such as CeRuSiH \cite{tence2008modulated}, LaFeSiH \cite{bernardini2018iron}, NdMnSiH \cite{tence2009huge}, and CeCoSiH/CeCoGeD \cite{chevalier2005influence}. Generally, upon hydrogenation of CeFeSi-type intermetallics via the solid-gas route, we observe a complete filling of the hydrogen site.  Therefore, the hydrides adopt the ZrCuSiAs structure as the 1111 iron-based superconductors \cite{kamihara2008iron,kamihara2006iron}. Details of PXRD refinements are available in supplementary material \cite{Supplementary}.

\begin{figure*}
    \centering
    \includegraphics[width=\linewidth]{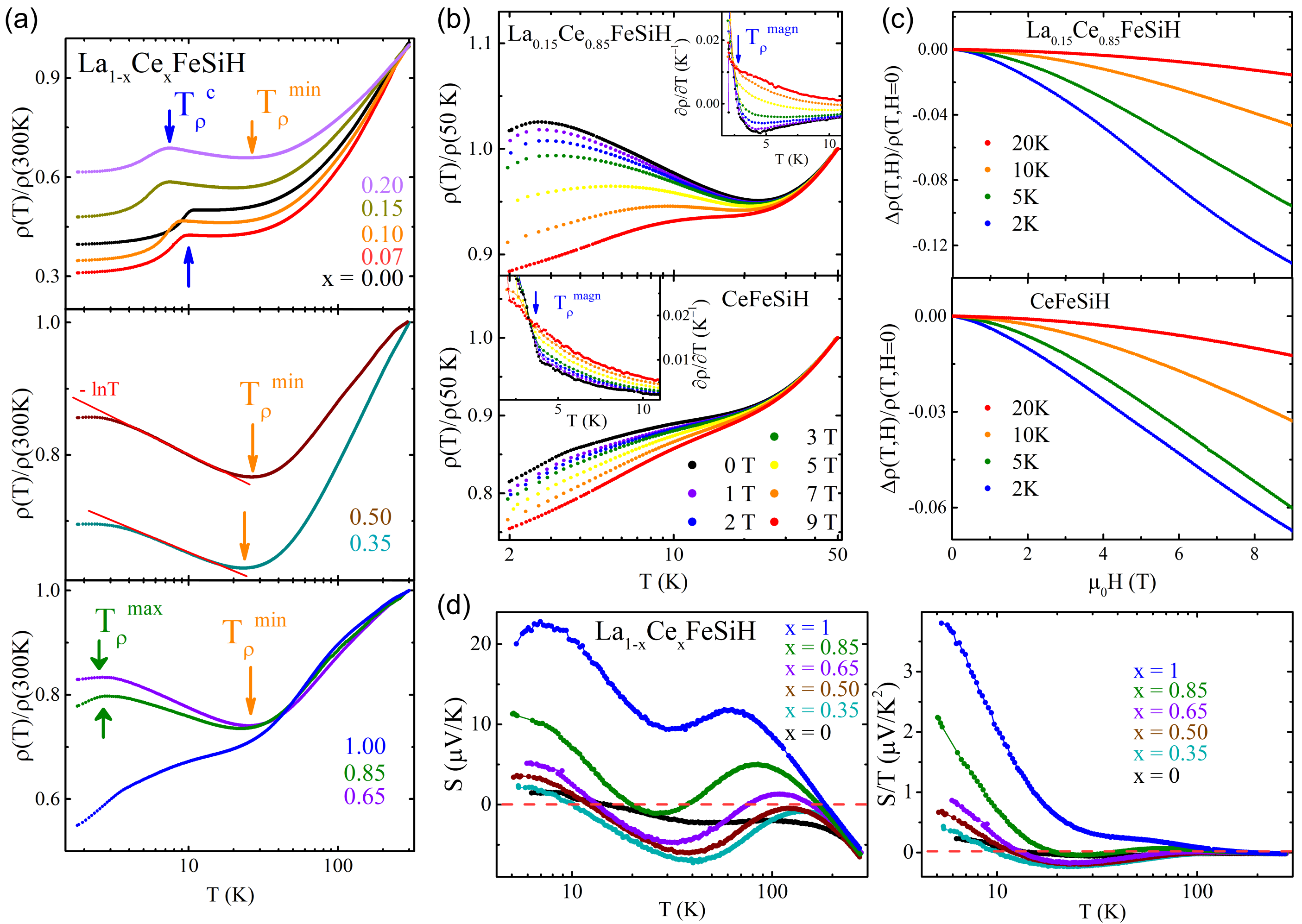}
    \caption{(a) Electrical resistance normalized at 300 K for the La$_{1-x}$Ce$_x$FeSiH solid solution as a function of temperature. (b) Electrical resistance normalized at 50 K for La$_{0.15}$Ce$_{0.85}$FeSiH and CeFeSiH. The insets show the derivative of the electrical resistance. (c) Magnetoresistance of La$_{0.15}$Ce$_{0.85}$FeSiH and CeFeSiH at low temperatures. (d) Thermopower for the La$_{1-x}$Ce$_x$FeSiH solid solution as a function of temperature.}\label{fig2} 
\end{figure*}
\

In Fig. \hyperref[fig1]{\ref*{fig1} (a)}, we compare the different interatomic distances between LaFeSiH and CeFeSiH. We note a contraction of 2.8$\%$ on the $c$ parameter, which goes from $c$ = 8.039 $\textup{\AA}$ in LaFeSiH to $c$ = 7.820 $\textup{\AA}$ in CeFeSiH, and a contraction of 0.8$\%$ on the $a$ parameter, which goes from a = 4.027 Å in LaFeSiH to $a$ = 3.996 $\textup{\AA}$ in CeFeSiH. Taking the metallic radius of cerium $r_{\text{Ce}}$ = 183 pm and of lanthanum $r_{\text{La}}$ = 188 pm from Ref. \onlinecite{teatum1968compilation}, we expect a contraction of 2.7$\%$. Thus, the observed variation of the $c$ parameter agrees with the expected metallic radius variation, which indicates an absence of variation for the oxidation state of the $R$H$^{x+}$ layer ($R$ = La, Ce). On the other side, the small contraction of 0.8$\%$ indicates that the a parameter is mostly imposed by the FeSi$^{x-}$ layer.  In particular, the Fe-Fe first neighbor distance variation is 0.8$\%$, going from d$^1_{\text{Fe-Fe}}$ = 2.848 $\textup{\AA}$ in LaFeSiH to d$^1_{\text{Fe-Fe}}$ = 2.826 $\textup{\AA}$ in CeFeSiH, and the Fe-Si first neighbor distance variation is 0.3$\%$, going from d$^1_{\text{Fe-Si}}$ = 2.357 $\textup{\AA}$ in LaFeSiH to d$^1_{\text{Fe-Si}}$ = 2.351 $\textup{\AA}$ in CeFeSiH. Thus, we expect that the $d$ electron physics is primarily unaffected by the substitution of lanthanum with cerium in the La$_{1-x}$Ce$_x$FeSiH system.

For the La$_{1-x}$Ce$_x$FeSiH solid solution, we observe a smooth decrease of both lattice parameters, as shown in Fig. \hyperref[fig1]{\ref*{fig1} (b)}. Thus, there is no evidence of any structural transition in this system, and we assume that the lanthanum-cerium substitution can be accounted for as only a change of the lanthanum $4f^0$ electronic configuration to the cerium localized $4f^1$ electronic configuration.

\subsection{Physical properties}

\subsubsection{Electrical resistance and thermoelectric power}

We present the electrical resistance measurements of the La$_{1-x}$Ce$_x$FeSiH hydrides as a function of temperature in Fig. \hyperref[fig2]{\ref*{fig2} (a)}. We observe three distinct behaviors as a function of the cerium concentration $x$. First, at a low cerium concentration $x \leq 0.2$, we observe the superconducting transition at a temperature $T^\rho_c$. The electrical resistance does not drop to zero even for LaFeSiH, but this is an artifact associated with grain boundary resistance in the compacted pellet used for the measurement since the resistivity reaches zero on LaFeSiH single crystal \cite{bernardini2018iron}. We justify the presence of superconductivity when it is coupled to a diamagnetic response of the magnetization, presented in a following section, even though we were not able to detect a diamagnetic signal in SQUID measurements for $x= 0.15$ and 0.20.

Secondly, for $0.07 \leq x \leq 0.85$, we observe a logarithmic increase of the electrical resistance below $T_\rho^{min}$, which is typical of the single-ion Kondo effect. We show this in detail in Fig. S2 of the supplementary material \cite{Supplementary}. For $x \geq 0.35$, the electrical resistance also reaches a maximum at very low temperatures, which indicates a many-ion Kondo-coherent regime below $T_\rho^{max}$. This low temperature maximum is quite similar to the example of the La$_{1-x}$Ce$_x$Cu$_2$Ge$_2$ system \cite{hodovanets2015remarkably}, which shows Kondo coherence even in the presence of an antiferromagnetic ground state. Finally, for CeFeSiH we do not observe anymore the logarithmic increase of the electrical resistance. We report the different temperature scales for the electrical measurements in Table \ref{table1}. Remarkably, the temperature scale for the Kondo-coherent regime $T_\rho^{max} \approx 2.5$ K is one order of magnitude smaller than for the Kondo-incoherent regime $T_\rho^{min} \approx 20$ K.

\begin{table}[h!]
    \centering
     \caption{Characteristic temperatures obtained on the La$_{1-x}$Ce$_x$FeSiH solid solution in electrical resistance measurements.}
     \label{table1}
     \scalebox{1}{\begin{tabular}{cccccccccc} 
        \toprule%
        \midrule
       $x$ & 0.00 & 0.07 & 0.10 & 0.15 & 0.20 & 0.35 & 0.50 & 0.65 & 0.85 \\
         \midrule
        $T_\rho^c$ (K) & 9.3 & 8.1  &  7.4 &  5.4 &  5.4 & -    & - & - & - \\
        $T_\rho^{min}$ (K) & -   & 15.3 & 17.1 & 20.1 & 24.3 & 23.5 & 26.5 & 25.5 & 22.5 \\
        $T_\rho^{max}$ (K) & -   & - & - & - & - & 2.25 & 2.25 & 2.7 & 2.85 \\
         \midrule
         \bottomrule
     \end{tabular}}
\end{table}

For the compounds with high cerium concentrations $x = 0.85$ and $x = 1$,  we present the electrical resistance for different values of the magnetic field in Fig. \hyperref[fig2]{\ref*{fig2} (b)}. We remark that the logarithmic increase of the resistance for La$_{0.15}$Ce$_{0.85}$FeSiH is strongly suppressed by the field, which suggests a magnetic origin. We observe a similar effect but smaller in magnitude for CeFeSiH. Furthermore, by inspecting the temperature derivative of the electrical resistance, we observe a strong change of slope at $T_\rho^{magn}$ = 2.8 K for $x = 0.85$ and  $T_\rho^{magn}$ = 3.8 K for $x = 1$. We have shown in a previous study that CeFeSiH has a transition towards a magnetic order at this temperature \cite{sourd2024drastic}. For $x = 0.8$5, $T_\rho^{magn}$ is slightly smaller than $T_\rho^{max}$, the temperature scale for which we expect Kondo coherence effects. Thus, a magnetic order may also be formed for $x = 0.85$ below $T_\rho^{magn}$, inside the Kondo-coherent Fermi liquid phase.
In order to investigate further the effect of the magnetic field on the electrical resistance, we present the magnetoresistance curves at low temperatures in Fig. \hyperref[fig2]{\ref*{fig2} (c)}. For both compounds, we observe a negative magnetoresistance, which cannot be explained by conduction electrons alone. As shown in Fig. S3 of the supplementary material \cite{Supplementary} at low fields the magnetoresistance is proportional to $H^2$. This behavior is expected for the single-ion Kondo effect \cite{hewson1993kondo}, and also observed several heavy-fermion systems like in CeCu$_6$ \cite{onuki1984magnetic} and in La$_{1-x}$Ce$_x$PtIn \cite{ragel2009dilution}.

\

We show our thermoelectric power $S(T)$ measurements as a function of temperature in Fig. \hyperref[fig2]{\ref*{fig2} (d)}. At high temperatures all the curves merge to a $T$-linear dependence, in accordance with the high-temperature limit of Kondo alloys, where the localized $4f^1$ electrons do not participate in the transport properties. At lower temperatures, we observe a very strong Seebeck coefficient in the compounds with high cerium concentrations. We observe a local maximum at about 100 K, which might be related to crystal-field effects as discussed in Ref. \onlinecite{sourd2024drastic}. While we observe a similar trend for the compositions $x = 0.35$, $x = 0.50$, and $x = 0.65$, the $x = 0.85$ and $x = 1$ compounds display a much higher thermopower at low temperatures. In a Fermi liquid picture, at low temperatures the thermopower should be proportional to the temperature, and the value of $S/T$ reflects the effective mass of the quasiparticles \cite{behnia2004thermoelectricity}. Thus, the much higher values of $S/T$ observed at low temperatures for $x = 0.85$ and $x = 1$ might indicate a heavy-fermion character of the quasiparticles, which is consistent with the Kondo coherence scenario suggested by the electrical resistance measurements.

\subsubsection{Magnetization and specific heat}

\begin{figure*}
    \centering
    \includegraphics[width=\linewidth]{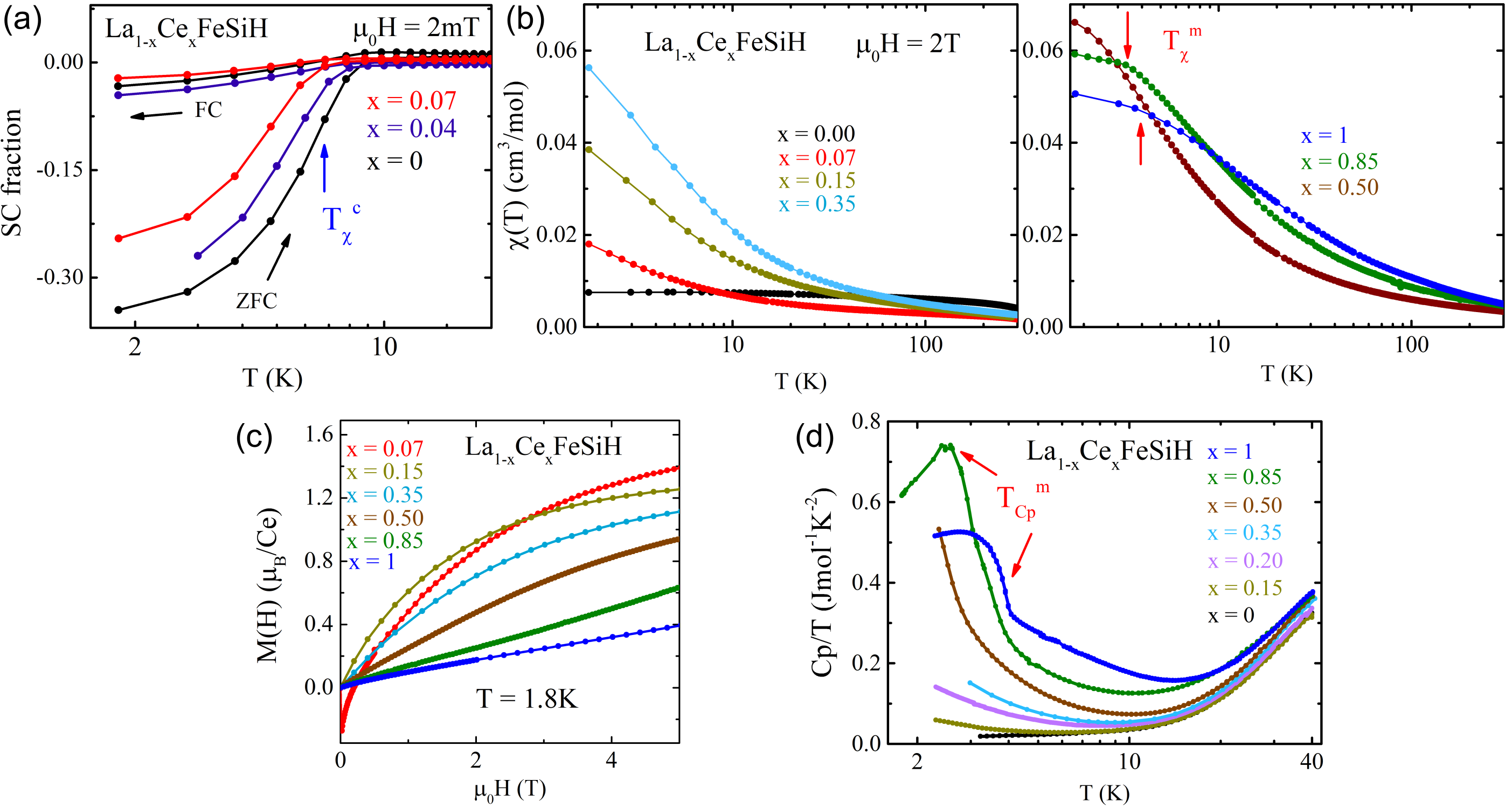}
    \caption{(a) Superconducting volume fraction of La$_{1-x}$Ce$_x$FeSiH compounds as a function of temperature for an applied magnetic field of 2 mT in zero-field cooled and field-cooled modes. We used a demagnetization factor of $N = 1/3$. (b) Magnetization of the La$_{1-x}$Ce$_x$FeSiH solid solution as a function of temperature for an applied magnetic field of 2 T. (c) Magnetization of the La$_{1-x}$Ce$_x$FeSiH solid solution as a function of magnetic field. (d) Temperature dependence of the specific heat divided by temperature for the La$_{1-x}$Ce$_x$FeSiH solid solution.}\label{fig3} 
\end{figure*}

In magnetization measurements, two different behaviors are evidenced. At low cerium concentration $x \leq 0.07,$ we observed Meissner effect which confirms the presence of superconductivity, as shown in Fig. \hyperref[fig3]{\ref*{fig3} (a)}. The critical temperature below which the Meissner effect appears, $T_\chi^c$, can be extracted, and this temperature decreases when adding cerium to the system. We show the magnetization versus field in the superconducting regime in  Fig. S4 of the supplementary material \cite{Supplementary}.

\

At higher cerium concentrations, the magnetism is dominated by the cerium contribution, as shown in Fig. \hyperref[fig3]{\ref*{fig3} (b)}. At high temperature, the magnetization curve of CeFeSiH follows a Curie-Weiss behavior with a magnetic moment of 2.48$\pm 0.1$ $\mu_B$ as shown in \onlinecite{sourd2024drastic}. At low temperature, the cerium contribution exhibits two different regimes: first, for $x = 0.07$, $0.15$, and $0.35$, we observe a non-saturating magnetization down to low temperatures. On the other hand, for the highly concentrated compounds, the magnetic susceptibility is nearly saturated at low temperatures. In particular, for $x = 0.85$ we observe a clear kink in the magnetic susceptibility, which suggests a magnetic transition at $T_\chi^m$ = 3.0 K. It is noteworthy that for $x = 1$, we have previously reported a magnetic transition at $T_\chi^m$ = 3.5 K, which is better visualized at lower applied magnetic fields as discussed in Ref. \onlinecite{sourd2024drastic}.This low transition temperature is typical in iron-based cerium systems such as CeFeAsO with $T_{Ce}^{m} = 3.7 K$ \cite{jesche2009rare}. Finally, we summarize the different temperature scales obtained from magnetization in Table \ref{table2}.

\begin{table}[h!]
    \centering
     \caption{Characteristic temperatures obtained on the La$_{1-x}$Ce$_x$FeSiH solid solution in magnetization measurements. $T_\chi^c$ denotes the temperature below which we observed the Meissner effect, and  $T_\chi^m$  the temperature where we observe a kink of the $M(T)$ curve at 2 T, indicating magnetic ordering.}
     \label{table2}
     \scalebox{1}{\begin{tabular}{cccccc} 
        \toprule%
        \midrule
       $x$ & 0.00 & 0.07 & 0.15 & 0.85 & 1.00 \\
         \midrule
        $T_\chi^c$ (K) & 8.0 & 7.0 &  6.0 &  - &  -  \\
        $T_\chi^{m}$ (K) & -   & - & - & 3.0 & 3.5  \\
         \midrule
         \bottomrule
     \end{tabular}}
\end{table}

Furthermore, at high cerium concentration we note a decrease of the low-temperature magnetic susceptibility as $x$ increases, which is clearly visible for $x \geq 0.50$. In addition, we show in Fig. S5 of the supplementary materials that the magnetization per cerium is a decreasing function of the cerium concentration in the whole temperature range. A similar evolution has been reported for La$_{1-x}$Ce$_x$Ni$_2$Ge$_2$ \cite{pikul2010lack}, where the magnetization per cerium of La$_{0.9}$Ce$_{0.1}$Ni$_2$Ge$_2$ is about 7 times higher than the one of CeNi$_2$Ge$_2$. The same trend is also observed in the La$_{1-x}$Ce$_x$Fe$_2$Ge$_2$ solid solution \cite{sugawara2000metamagnetic}. This progressive reduction of the magnetic moment per cerium is justified in Ref. \onlinecite{pikul2010lack} as the manifestation of the Kondo effect, leading to a progressive screening of the cerium magnetic moment. A second possible scenario is a size effect, due to the larger radius of the lanthanum atom with respect to the cerium atom. In the Doniach picture \cite{doniach1977kondo}, the substitution with a larger atom has the effect of a negative pressure, which favors the localized magnetic character of the cerium $4f^1$ electron, while the substitution with a smaller atom favors the itinerant character of the $4f^1$ electron. This effect has been studied in detail in the La$_{1-x}$Ce$_x$Ru$_2$Si$_2$ and Y$_{1-y}$Ce$_y$Ru$_2$Si$_2$ systems \cite{haen1988magnetization}, where the non-magnetic yttrium and lanthanum atoms are, respectively, smaller and larger than the cerium atom. The authors of Ref. \onlinecite{haen1988magnetization} demonstrated that the magnetization per cerium atom increases upon decreasing $x$, while it decreases upon decreasing $y$.

We present magnetization curves versus field at 1.8 K in \hyperref[fig3]{\ref*{fig3} (c)}. For $x = 0.07$, we observe a negative magnetization at low field due to the superconducting regime, which is rapidly suppressed at higher field. For increasing $x$, we observe a decrease of the magnetization per cerium atom, in accordance with the temperature dependence (see Fig. S5 in \cite{Supplementary}). Furthermore, we observe that the shape of the magnetization curve evolves from a Brillouin-like function for $x = 0.15$ to a straight line for $x = 1$. This could be associated with the progressive increase of antiferromagnetic correlations in this system at low temperatures.

\

We show the specific heat divided by the temperature as a function of temperature in \hyperref[fig3]{\ref*{fig3} (d)}. The measurement on LaFeSiH reveals a very weak contribution, which does not permit resolving the superconducting transition. We show this in detail in Fig. S6 of the supplementary material \cite{Supplementary}. From the value of $Cp/T$ at the lowest temperature measured, we estimate the Sommerfeld coefficient of LaFeSiH at 19 mJ/ (mol K$^2$) which is 27 times less than the one reported for CeFeSiH in \onlinecite{sourd2024drastic}. Upon increasing the cerium concentration $x$, we observe an increase of the specific heat at low temperatures. For $x = 0.85$, we detect a transition as a peak of the $Cp/T$ curve at $T_{Cp}^m$ = 2.5 K. For CeFeSiH, the transition is associated with a jump of $Cp/T$ at $T_{Cp}^m$ = 3.5 K. Together with the magnetization anomaly, this suggests the establishment of a magnetic order for these compounds at low temperatures. Finally, for La$_{0.50}$Ce$_{0.50}$FeSiH, we observe a diverging $Cp/T$ down to 2.5 K. This could correspond to a quantum-critical point or to a magnetic transition occurring at a lower temperature. A third possibility could be the presence of a spin-glass state, as recently proposed for the closely related La$_{1-x}$Ce$_x$FePO solid solution \cite{chen2018spin}. While magnetic susceptibility measurements would allow a more direct evidence of glassy behavior, we do not expect our specific heat results to distinguish between spin glass and antiferromagnetic order \cite{mydosh1978spin}. 

% Considering the negative pressure effect induced by the La atoms, the realization of magnetic order can be understood from the Doniach argument \cite{doniach1977kondo}.

\subsection{Phase diagram and discussion}

\begin{figure}
    \centering
    \includegraphics[width=\linewidth]{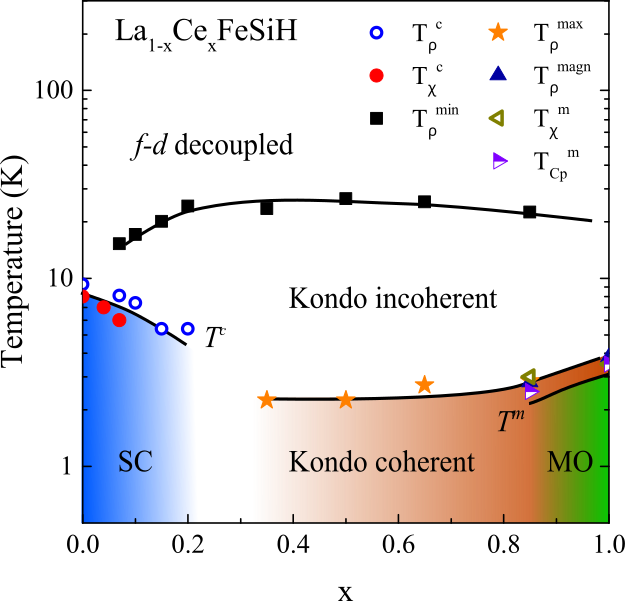}
    \caption{ Phase diagram representing the different regimes observed in the La$_{1-x}$Ce$_x$FeSiH solid solution. SC stands for superconducting, and MO for magnetic order. The separation between $T_\rho^{max}$ and $T^m$ has been magnified for clarity.}\label{fig4} 
\end{figure}

We gather the different temperature scales extracted from the various transport and thermodynamic probes in order to build the phase diagram of Fig. \hyperref[fig4]{\ref*{fig4}}. This phase diagram shows three different regimes. At low temperatures and low cerium concentration $x$, we observe superconductivity in electrical resistance and magnetization measurements. 

 For a wide range of cerium concentrations, $0.07 \leq x \leq 0.85$, we detect signatures of the Kondo incoherent regime in electrical resistance for $T<T_\rho^{min}$. For medium-large cerium concentration $0.35 \leq x \leq 0.85$, below Tmin we could identify also a Kondo coherent regime for $T< T_\rho^{max}$ . For $x > 0.85$ a magnetically ordered phase could be put in evidence by specific heat. 

% For a wide range of cerium concentrations, $0.07 \leq x \leq 0.85$, we detect signatures of the Kondo incoherent regime in electrical resistance. Finally, for large cerium concentration $x \geq 0.35$, we identify the Kondo coherent regime coexisting with a possible magnetically ordered phase at low temperatures for $x > 0.85$. 

At high temperatures, the iron $d$ electrons and the cerium $f$ electrons are decoupled, and the La$_{1-x}$Ce$_x$FeSiH systems behave like a normal metal with localized magnetic moments. This can be deduced from the thermoelectric power or the specific-heat curves in Figs. \hyperref[fig2]{\ref*{fig2} (d)} and \hyperref[fig3]{\ref*{fig3} (d)}, which are virtually independent of the cerium concentration in the high-temperature limit. Below $T_\rho^{min}$, the $d$ and $f$ electrons start to interact, leading to the logarithmic increase of the electrical resistance at low temperatures, typical of the Kondo effect. Finally, at low temperatures, the electronic correlations drive the system into different quantum phases that can be tuned by the cerium concentration.

\

On the one hand, the iron-based superconductivity of LaFeSiH is well established from from magnetization (with almost 100$\%$ of SC volume) \cite{bernardini2018iron}, band-structure calculations \cite{bernardini2018iron}, magnetic penetration depth \cite{bhattacharyya2020evidence}, and nuclear magnetic resonance data \cite{hansen2024magnetic}.  In the La$_{1-x}$Ce$_x$FeSiH system, we observe the superconducting regime below $x = 0.07$ from our magnetization measurements and below $x = 0.20$ from our electrical resistance measurements.  Remarkably, in this concentration range $0.07 < x < 0.20$, we also observe Kondo effect with an increase of the electrical resistance below $T_\rho^{min}$. Thus, in this range of cerium concentration, the effects of $f$ electron correlations are not able to destroy the superconductivity at low temperatures. Indeed, the decreasing of the superconducting transition temperature with $x$ could correspond to a pressure effect induced by the smaller atomic radius of cerium, since it has been shown that a pressure of a few GPa is enough to strongly reduce $T_c$ on LaFeSiH single crystals \cite{bernardini2018iron}.

\

%  The situation is less clear for $x = 0.5$, where the magnetic susceptibility shows a small saturation at low temperatures, but no saturation is observed on the $Cp/T$.

On the other hand, the magnetic order of $f$ electrons is well established above $x = 0.85$, with the presence of specific heat and magnetization anomalies. For $x = 0.5$ and $x = 0.35$, we observe a low temperature maximum of the electrical resistance, which does not permit us to affirm the existence of a magnetic order at lower temperatures, as paramagnetic Kondo coherence effects might also create a comparable signature. The clear establishment of a Kondo coherent regime for $0.5 \leq x \leq 1$ would require experimental data at lower temperatures, as for instance in the case of the La$_{1-x}$Ce$_x$Cu$_6$ solid solution, where the typical $T^2$ dependence of the resistivity is observed only below 0.1 K \cite{sumiyama1986coherent}.

% Nevertheless, our observations of a maximum in electrical resistance, a saturation of the magnetic susceptibility, and a large value of the specific heat at low temperatures make this proposition quite plausible. 

Thus, upon varying the cerium concentration $x$, we are able to change between a superconducting ground state and a Kondo coherent ground state at low temperatures. Our observation of a maximum of electrical resistance for $x\geq 0.35$ is consistent with the scenario of a Kondo coherent-Kondo dilute transition proposed in Refs. \cite{poudel2021phase,burdin2007random}. Furthermore, approaching the transition, the superconducting critical temperature $T^c$ as well as the Kondo coherent and magnetic order temperatures $T_\rho^{max}$ and $T^m$ decrease and go below our experimental range. This could suggest the presence of a quantum-critical point, with non-saturating magnetic susceptibility at low temperatures as for La$_{0.65}$Ce$_{0.35}$FeSiH [Fig. \hyperref[fig3]{\ref*{fig3} (b)}] and non-saturating $Cp/T$ as for La$_{0.5}$Ce$_{0.5}$FeSiH [Fig. \hyperref[fig3]{\ref*{fig3} (d)}]. A second possible explanation could be the presence of a magnetic ordering at lower temperatures.

\section{Conclusion}

We realized the synthesis and the characterization of the La$_{1-x}$Ce$_x$FeSiH solid solution. All the compounds crystallize in the same space group $P4/nmm$. Physical measurements evidence complex behavior such as superconductivity, Kondo effect, and magnetic order.

 In order to characterize the physical properties of the La$_{1-x}$Ce$_x$FeSiH solid solution, we applied a systematic study methodology that allows extracting temperature scales for all the different regimes observed. We evaluate the critical temperature for the superconducting transition in electrical resistance $T_\rho^c$ and in magnetization $T_\chi^c$, a temperature for incoherent Kondo regime $T_\rho^{min}$, and temperatures associated with Kondo coherence and magnetic order in electrical resistance $T_\rho^{max}$ and $T_\rho^{magn}$, in magnetic susceptibility $T_\chi^{m}$, and in specific heat $T_{Cp}^{m}$. Thanks to these different energy scales, we build a phase diagram that shows the evolution from the superconducting regime of LaFeSiH and the Kondo coherent regime of CeFeSiH.

Our study shows how the superconducting regime of LaFeSiH, which is generated by correlated $d$ electrons, is suppressed in favor of the Kondo coherent regime of CeFeSiH, which is generated by correlated $f$ electrons. In particular, we reveal that the transition from one regime to another is particularly complex. At low cerium concentrations, we observed the coexistence of Kondo effect and superconductivity at $0.07 < x < 0.20$ indicating the presence of correlated $d$ and $f$ electrons. At higher cerium concentrations $0.25 < x < 0.85$, we detected signatures of Kondo coherence but no phase transition down to 2 K, which could correspond to a putative quantum-phase transition arising from an exotic interplay of correlated $d$ and $f$ electrons. Finally, at $x = 0.85$, we observed coherent Kondo lattice effects coexisting with a magnetic ordered phase at low temperatures. For $x=1$ only a clear evidence of a magnetic ordered phase could be put in evidence.

% Finally, at the highest cerium concentration $x > 0.85$, we observed coherent Kondo lattice effects and a magnetic order at low temperatures.

In short, we demonstrated that the cerium concentration $x$ in the La$_{1-x}$Ce$_x$FeSiH system serves as a unique control parameter, bridging the correlated $d$ electron physics of iron-based superconductors with the correlated $f$ electron physics of heavy fermions. To further explore the change of ground state around $x = 0.3$, the study of single crystals and access to lower temperatures will be invaluable.

\section{Aknowledgments}
We thank Bernard Chevalier and Jean-Baptiste Vaney for fruitful discussions. J. Sourd acknowledge the support of the High Magnetic Field Laboratory (HLD) at Helmholtz-Zentrum Dresden-Rossendorf (HZDR), a member of the European Magnetic Field Laboratory (EMFL), the Deutsche Forschungsgemeinschaft (DFG) through SFB 1143, and the Würzburg-Dresden Cluster of Excellence on Complexity and Topology in Quantum Matter ct.qmat (EXC 2147, Project No. 390858490).

\bibliography{biblio_LaCeFeSiH.bib}

\end{document}